\documentclass[aps, prl, reprint, amsfonts, amssymb, amsmath, preprintnumbers, showpacs, nofootinbib,superscriptaddress]{revtex4-1}
\usepackage{graphicx, amsthm, multirow}
\usepackage[citecolor=blue]{hyperref}
\usepackage[all]{hypcap}
\usepackage{url}
\usepackage{feynmp-auto}
\usepackage{xcolor}
\usepackage{subcaption}
\usepackage{caption}
\def\lagrangian{lagrangian}
\def\eg{\emph{e.g.}}

\begin{document}

\title{U$(1)^\prime$ mediated decays of heavy sterile neutrinos in MiniBooNE}

\author{Peter Ballett}
\email{peter.ballett@durham.ac.uk}
\author{Silvia Pascoli}
\email{silvia.pascoli@durham.ac.uk}
\affiliation{Institute for Particle Physics Phenomenology, Durham University, Durham DH1 3LE, UK.}
\author{Mark Ross-Lonergan}
\email{markrl@nevis.columbia.edu}
\affiliation{Columbia University, Nevis Laboratories, New York, NY, 10027, USA.}

\date{\today}

\begin{abstract}

The MiniBooNE low-energy excess is a longstanding problem which has received further confirmation recently with a reanalysis using newly collected data, with the anomaly now at the $4.8\sigma$ level. In this letter we propose a novel explanation which advocates a low-energy sector containing $Z^\prime$ bosons with GeV-scale masses and sterile neutrinos with masses around $100$--$500$ MeV. We show that this scenario can provide an excellent spectral agreement with the MiniBooNE low-energy excess in the form of $Z^\prime$-mediated neutral current production of heavy sterile states, a fraction of whose subsequent decay to $e^+e^-$ pairs are misidentified as single electron-like electromagnetic showers. Our model inscribes itself in the broad class of models in which sterile neutrinos are charged under novel interactions, allowing new couplings to hidden-sector physics. Alongside the electron-like MiniBooNE signature this model also predicts a novel, low-background, signal in LArTPC detectors such as MicroBooNE consisting of two distinguishable electron-like electromagnetic showers originating from a single vertex with no associated hadronic activity.  

\end{abstract}

\preprint{IPPP/18/70} 

\maketitle

\section{Introduction}
The MiniBooNE low-energy excess (LEE) \cite{AguilarArevalo:2008rc} is a longstanding anomaly which has received renewed attention thanks to a recent analysis~\cite{Aguilar-Arevalo:2018gpe}, which doubled the amount of data in neutrino mode sample, confirming the existence of the LEE at the $4.8\sigma$ level. 
The excess of approximately 460 low-energy electron-like ($e$-like) events was observed in both neutrino and, to a lesser extent, anti-neutrino channels with $12.84 \times 10^{20}$ protons on target with the Booster beam in neutrino mode, and $11.27 \times 10^{20}$ protons on target in antineutrino mode \cite{Aguilar-Arevalo:2018gpe}.
The events are almost entirely contained within the lowest energy bins of the
experiment, $E \lesssim 0.6$ GeV visible energy, and their angular distribution is relatively flat, with a slight preference to being forward \cite{Aguilar-Arevalo:2018gpe}.

These events have yet to receive a satisfactory explanation, be it through conventional or unconventional physical mechanisms. 
One of the most popular interpretations of this signal is via the oscillation
$\nu_\mu\to\nu_e$. This would point to an additional oscillatory
frequency, driven by a mass-squared splitting $ \Delta	m^2 \approx
\mathcal{O}(1)$~eV$^2$, requiring the existence of a sterile neutrino.
This solution furthered the intrigue around a collection of anomalous short-baseline oscillation results pointing to possibly related effects \cite{PhysRevLett.77.3082,
PhysRevLett.75.2650}. The LEE spectrum agrees with that of an oscillatory solution \cite{Aguilar-Arevalo:2018gpe} and is consistent with the LSND anomaly; however, such an explanation remains controversial for two main reasons.
Firstly, global fits of the oscillation data results show significant tension between experiments. Appearance experiments such as LSND and MiniBooNE require relatively large mixing angles between the nearly-sterile neutrino and both the electron and muon neutrinos. The electron reactor neutrino experiments NEOS~\cite{PhysRevLett.118.121802} and DANSS~\cite{Alekseev:2018efk} favour a sterile neutrino oscillation explanation with a mixing with the electron neutrino around $U_{e4} \sim 0.1$, while $\nu_\mu$ disappearance experiments, driven mainly by MINOS+ and IceCube, strongly constrain the $U_{\mu 4}$ mixing element, naively excluding this explanation of the LEE. See \cite{Dentler:2018sju},\cite{Gariazzo:2017fdh} and \cite{Collin:2016aqd} for recent overviews of the global situation. The upcoming Fermilab Short Baseline Neutrino program aims to definitively confirm of refute this hypothesis \cite{Antonello:2015lea,Cianci:2017okw}. Secondly, a nearly-sterile neutrino with the masses and mixing required to explain the LEE would pose problems for our understanding of the early Universe. A light sterile would thermalize in the Early Universe providing an extra degree of freedom at the time of Big Bang Nucleosynthesis and the Cosmic Microwave Background as well as producing an overly large contribution to hot dark matter. Some solutions have been proposed for this problem, by considering a non-minimal model where the sterile neutrinos also feel a new interaction. In these so called ``secret-interaction'' scenarios, sterile neutrino production can be dynamically suppressed in the early Universe through finite-temperature effects \cite{Hannestad:2013ana,Dasgupta:2013zpn}. Although such models reduce cosmological tension, they do not completely remove it and the parameter space is being increasingly constrained by new cosmological data \cite{Chu:2018gxk,Song:2018zyl}.

Non-oscillatory explanations of the excess have also been suggested
\cite{Gninenko:2009ks, Gninenko:2010pr, Masip:2012ke} which typically postulate the production of heavy sterile neutrinos in scattering events \emph{inside}
the detector \cite{Gninenko:2009ks, Gninenko:2010pr}.  As a mineral oil \v{C}erenkov detector, MiniBooNE lacked any capability for separation of electrons and photons, both producing near identical \v{C}erenkov cones inside the detector \cite{AguilarArevalo:2008qa}. As such, the observed excess events are only definitively known to be electromagnetic (EM) showers, whose origins are unknown. These non-oscillatory solutions postulate that the signal is in fact made of single photons, which arise from the radiative decay of the heavy nearly-sterile neutrinos with masses of $\mathcal{O}(10\text{--}100)$ MeV. In the first version of this explanation \cite{Gninenko:2009ks, Gninenko:2010pr}, even for a large decay rate, the total number of signal events is severely bounded by the rate of heavy sterile neutrino production, which is mediated by Standard Model (SM) weak neutral-current (NC) and suppressed by sterile mixing angles. To get a sufficient number of events, very large mixing angles between the active and nearly-sterile neutrinos are required, $10^{-2} \lesssim |U_{\mu 4}|^2 \lesssim 10^{-3}$, which sit very uneasily with the bounds from prior experiments, and an extremely large transition magnetic moment. Subsequently, sterile masses below $100$ MeV were found to be in tension with radiative muon capture rates measured at TRIUMF \cite{McKeen:2010rx} and of rare Kaon decays by the ISTRA+ collaboration \cite{Duk:2011yv}.
A variant was proposed which evaded some of these constraints by using the neutrino-photon vertex to drive both the initial production and subsequent decay of the heavy states \cite{Masip:2012ke}. However, the photon-mediated scattering, which prefers small $Q^2$-exchange with the nucleus, results in events strongly clustered around the beamline \cite{Radionov:2013mca}, in contrast to the flatter angular distribution of the LEE.
Recent work on neutrino dipole portals have placed further constraints on the parameter space of magnetic moment explanations \cite{Magill:2018jla}.

In this letter, we discuss a novel explanation of the excess, based on the idea of heavy nearly-sterile neutrino production \emph{in situ}, that combines the kinematic behaviour of production by a heavy mediator with an enhanced decay rate. The core idea is that neutrinos have NC interactions mediated by a new GeV-scale boson such that heavy sterile neutrinos with masses $100 \ \mathrm{MeV}  \lesssim m_4 \lesssim 500$ MeV are produced by neutrino beam interactions with the nuclei in the detector. These subsequently decay into $e^+ e^-$ pairs giving rise to the signal through mis-reconstruction. 
The introduction of a new boson significantly enhances the production cross section if its mass is below 10 GeV, allowing for smaller values of $U_{\mu 4}$, while avoiding the problem encountered by explanations based on photon exchange of failing to well reproduce the angular spectrum \cite{Radionov:2013mca}.
Another crucial aspect of our work is the reinterpretation of the excess as an $e^+e^-$ pair. As we discuss, this can occur if either the electrons are closely overlapping or if one of the two fermions is too soft to be reconstructed as a shower. With this novel interpretation of the excess, we open up the possibility for a NC process driving both the production and decay of heavy sterile neutrinos inside the detector. 

For simplicity, we assume that the heavy sterile neutrino production and its enhanced decay rate come from just one novel interaction, which for definiteness we take to be a $U(1)^\prime$ kinetically mixed with hypercharge. This assumption is not essential to our proposal, indeed these two steps may easily be mediated by different bosons, but we leave the discussion of such variants to future work.


\section{\label{sec:model} The phenomenological model}

We assume that the SM gauge group is extended by a new factor U$(1)^\prime$
\cite{Fayet:1980ad}.  In the most general \lagrangian, the
gauge boson of this symmetry kinetically mixes with hypercharge through a mixing parameter
 $\chi$. We assume that the new gauge symmetry is broken at low
energies, and include a mass for the $X$ boson without detailing its provenance
or specifying its scale. The low-scale \lagrangian\ is then given by
\begin{align*} \mathcal{L} &= \mathcal{L}_{\nu\text{SM}}
-\frac{1}{4}X_{\mu\nu}X^{\mu\nu} - \frac{\sin\chi}{2}X_{\mu\nu}B^{\mu\nu} +
\frac{\mu^2}{2}X_\mu X^\mu, \end{align*}
where $\mathcal{L}_{\nu\text{SM}}$ denotes an extension of the SM incorporating
neutrino masses which we will return to later, $F_{\mu\nu}\equiv\partial_\mu
F_\nu - \partial_\mu F_\nu$ with $B_\mu$ and $X_\mu$ denoting the
U$(1)_\text{Y}$ and U(1)$^\prime$ gauge fields, respectively. 
As usual, the kinetic mixing term between $B_\mu$ and $X_\mu$ can be removed by
a field redefinition \cite{Langacker:2008yv,Rizzo:2006nw}, and we further identify the states of definite mass (denoted $A$, $Z$ and $Z^\prime$) by performing a change of basis between these fields and the third generator of the SU$(2)_\text{L}$ group. This
transformation can be expressed at first order in $\chi$ (and zeroth order in
$\mu/v$, with $v$ denoting the Higgs VEV) by
\begin{align*} \left (\begin{matrix} A \\ Z \\ Z^\prime \end{matrix} \right ) =
\left(\begin{matrix} c_\text{W} & s_\text{W} & c_\text{W}\chi\\ -s_\text{W}&
c_\text{W} & 0\\ s^2_\text{W}\chi & -s_\text{W} c_\text{W}\chi & 1
\end{matrix}\right) \left (\begin{matrix} B \\ W^3 \\ X \end{matrix}
\right),\end{align*}
where $s_\text{W} \equiv \sin\theta_\text{W}$ and $c_\text{W}\equiv
\cos\theta_\text{W}$ with $\tan\theta_\text{W}=g^\prime/g$.  After electroweak symmetry breaking, the mass of the
photon (denoted by $A$) vanishes exactly, while to first order in $\chi$, 
the $Z$ has the SM expression for its mass and the $Z^\prime$ has a mass 
given by $\mu$.

We assume that none of the SM field content is charged under the novel
$U(1)^\prime$. Working to first order in $\chi$ and $\mu/v$, the coupling between a SM fermion $f$ and the $Z^\prime$ is purely vectorial and proportional to both $\chi$ and the particle electric charge $q_f$,
\[  \mathcal{L}\supset -e q_f c_\text{W} \chi \overline{f}\gamma^\mu f Z^\prime_\mu. \]
Although this naively precludes neutrino-$Z^\prime$ interactions, we introduce SM-gauge 
singlets into $\mathcal{L}_{\nu\text{SM}}$ which are charged under the 
new U$(1)^\prime$, and assume that they mix with the SM neutrinos.
This scenario requires that the neutrino mass terms violate the novel symmetry, which is a rather generic feature if the theory contains new scalars which spontaneously break the $U(1)^\prime$.
In this letter, rather than present a complete model of the neutrino sector, we work with a simplified scheme which captures the essential phenomenology whilst retaining a significant degree of model-independence.
For simplicity, we work with a single right-handed neutrino, but the extension
to include multiple states is unproblematic and usually necessary in a fully consistent theory.
In fact, a single Majorana neutrino with the mass and mixing angles required to explain the MiniBooNE signal would typically lead to unacceptably large contributions to the light neutrino mass matrix. This issue can be solved in linear or extended see-saw models in which two sterile neutrinos are introduced: their contributions to neutrino masses nearly cancel out allowing for large mixing angles and small neutrino masses at the same time.
We denote the neutrino flavour states as $\nu_\alpha$
$\alpha=\{e,\mu,\tau,\text{s}\}$ and the mass eigenstates as $\nu_i$, $i=1,2,3,4$. These are related via a $4\times 4$  Pontecorvo-Maki-Nakagawa-Sakata(PMNS)-like matrix $U$ as $\nu_\alpha = U_{\alpha i} \nu_i$. We follow the convention that the mass 
index $j$ refers to the light states.
The nearly-sterile massive neutrino $\nu_4$ is assumed to have a small mixing with the lighter states, and will be referred to as a ``sterile neutrino".

Neutrino mixing mediates the U$(1)^\prime$ interaction to the light neutrinos, and we find the following neutrino-$Z^\prime$ vertices to leading-order in $\chi$ and the small elements of the PMNS-like matrix, 
\begin{align*}  \mathcal{L}\supset~ &~ U^*_{\alpha 4}g^\prime\overline{\nu_\alpha}\gamma^\mu P_\text{L}\nu_4Z^\prime_\mu + U^*_{\alpha 4}U_{\beta 4}g^\prime\overline{\nu_\alpha}\gamma^\mu P_\text{L}\nu_\beta Z^\prime_\mu\\
&+ g^\prime\overline{\nu_4}\gamma^\mu P_\text{L}\nu_4 Z^\prime_\mu, \end{align*}
where $g^\prime$ is the coupling constant of the new force. Provided that $|U_{\alpha 4}|^2 \ll 1 $, we can expect sterile-active-$Z^\prime$ interactions to occur at a higher rate than active-active-$Z^\prime$ interactions ensuring that SM processes are largely unaffected.
The novel interactions have a number of consequences for both sterile neutrino and $Z^\prime$ phenomenology. Firstly, we expect heavy neutrino production inside neutrino detectors as heavy neutrinos are produced via $Z^\prime$-mediated upscattering.
Secondly, these heavy states will have shorter lifetimes from enhanced NC decays. This could occur via either an on-shell or off-shell $Z^\prime$, \emph{e.g.} $\nu_4 \to \nu_\alpha Z^\prime$ or $\nu_4 \to \nu_\alpha e^+ e^-$, depending on the hierarchy of the $Z^\prime$ and heavy neutrino masses. Here, for definiteness, we focus on the case $m_4 < m_{Z^\prime}$. Although dependent on the precise values of kinetic and neutrino mixing parameters (and the possible presence of other particles in the model), order one branching fractions to $\nu_\alpha e^+e^-$ are attainable, with the other likely decays being to invisible multi-neutrino final states\footnote{Although kinematically possible for heavy neutrinos with masses above $135$ MeV, decays into pseudoscalar mesons are not enhanced by the $Z^\prime$ due to is vectorial coupling to quarks.}. 
Finally, for $m_{Z^\prime}>2m_4$ the $Z^\prime$ will have a dominant decay into two sterile neutrinos. Although the latter are unstable, this unconventional final state will impact the constraints from previous experiments.

The analysis in this paper will depend on both the production of a heavy neutrino via $Z^\prime$ mediated quasi-elastic scattering and the subsequent decay
of the heavy neutrino into an electron-positron pair and a light neutrino. 
We include both coherent scattering off of Carbon and incoherent scattering off the constituent protons of the detector medium. The coherent cross section is computed using an analytical approximation of a Woods-Saxon form factor \cite{Fricke:1995zz,Jentschura2009} based on the symmetrized Fermi function \cite{Anni1994,Sprung1997}. The hadronic current in the neutrino-proton cross section is parameterized by the electromagnetic form factors of the proton \cite{Formaggio:2013kya}, as the $Z^\prime$ only couples to SM particles via its electric charge. 

\section{The MiniBooNE low energy excess}
\begin{figure*}[t]
 \begin{center}
 \begin{subfigure}{0.45\textwidth}
        \centering
	 \includegraphics[width=\linewidth]{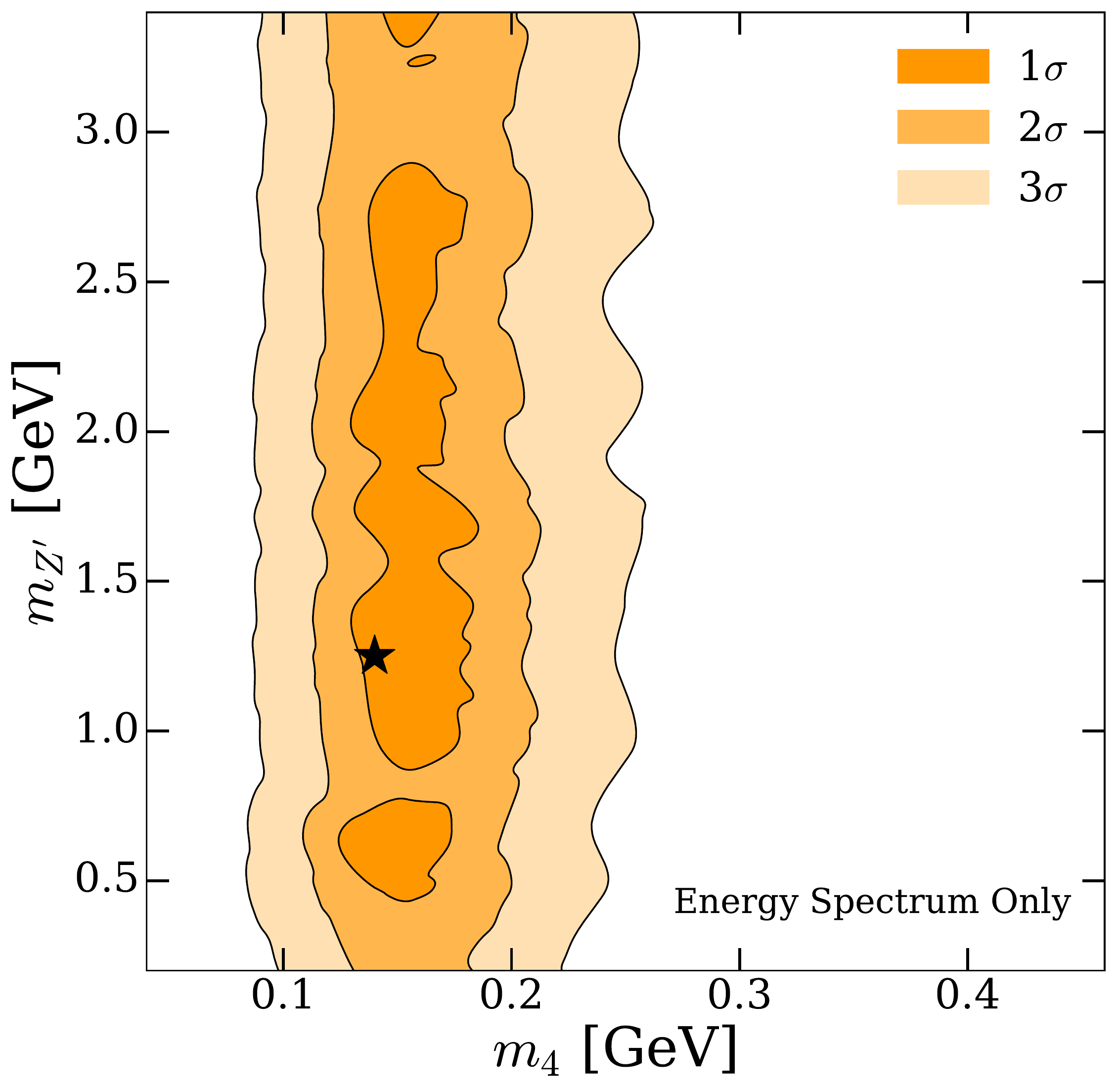}
    \end{subfigure}%
    ~ 
    \begin{subfigure}{0.45\textwidth}
        \centering
 \includegraphics[width=\linewidth]{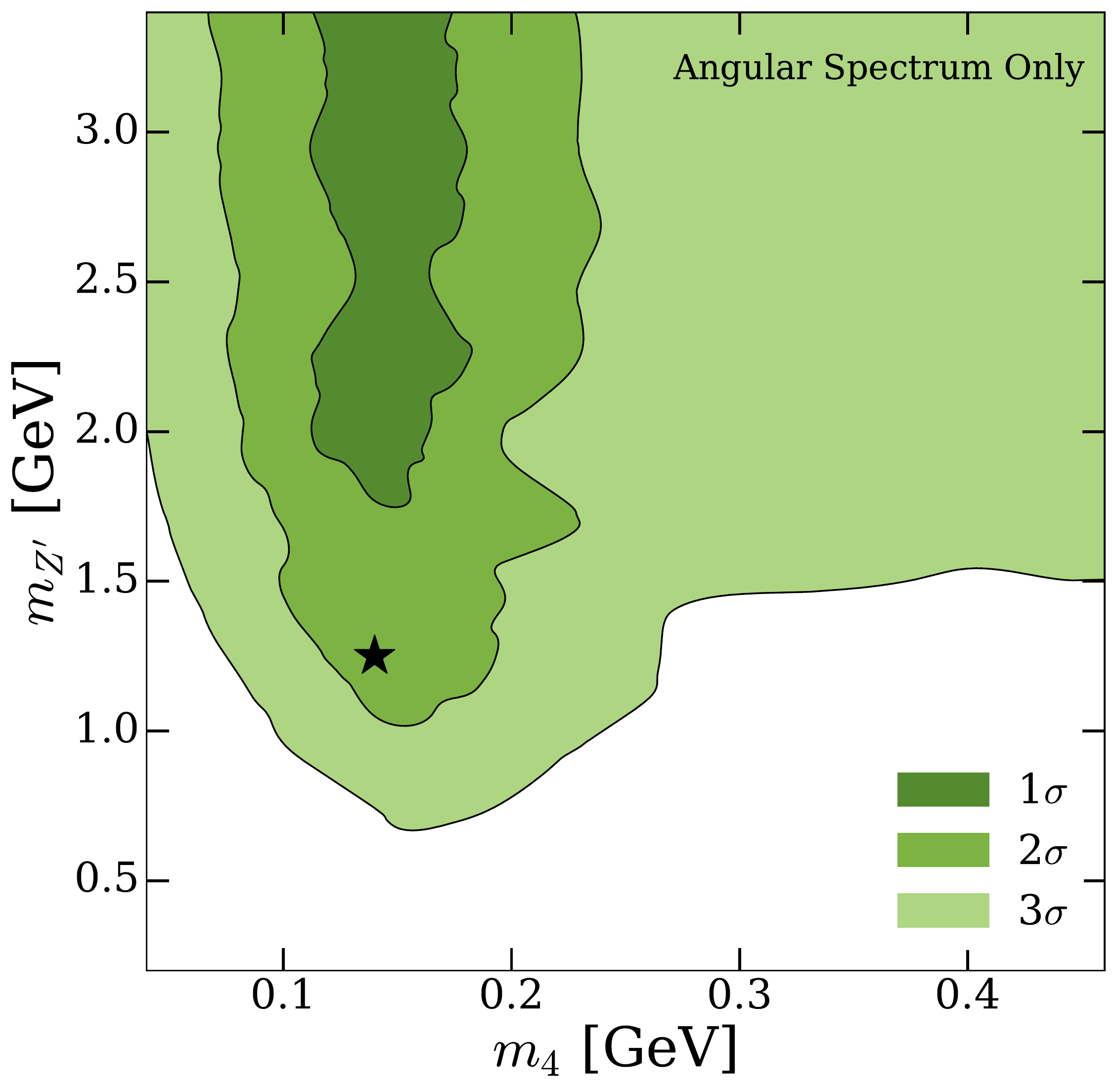}
    \end{subfigure}
\end{center}
\caption{\label{fig:ee_cuts} Results of parameter scan in $m_4$ and $m_{Z^\prime}$ fitting the shape of our model signal to reconstructed visible energy spectrum (left) and the reconstructed shower angle (right). The energy spectra is sensitivity only to $m_4$, requiring $ 100 \lesssim m_4 \lesssim 200$ MeV to produce an excess with low enough energy, where as the angular spectrum puts a lower bound on the $Z^\prime$ mass below which the signal events are too forward going.  The black star shows a representative point, detailed in Figure (\ref{fig:spectra}).}
\end{figure*} 

In the decay pipe of the Booster Neutrino beam \cite{AguilarArevalo:2008yp}, a flux of both light neutrinos and heavy neutrinos are incoherently produced in meson decays, due to the large difference in mass. The heavy neutrinos decay before reaching the detector and MiniBooNE sees a beam constituted by a coherent superposition of light neutrinos $\nu_j$. Heavy neutrinos are then re-produced inside the detector by the novel $Z^\prime$-mediated production process  $\nu_\mu + {\cal N} \to \nu_4 + {\cal N}$ for a target hadron ${\cal N}$.
For this to give a number of events comparable to few \% of the NC ones due to $\nu_\mu$ interactions, we require that $|U_{\mu 4}|^2 \chi^2\left(m_Z/m_{Z^\prime}\right)^4 \sim 0.01$, which suggests a scale for the new ${Z^\prime}$ of below 10 GeV. $\nu_4$ are also produced by SM $Z$-mediated processes, but these are subdominant as they are suppressed by both the sterile mixing angle and the heavy $Z$ mass. 
In our model, the subsequent decay of these heavy neutrinos produces the MiniBooNE excess events.
The dominant visible decay rate is to $e^+ e^-$ pairs mediated by the new boson
\[ \nu_4 \to \nu_\alpha~e^+~e^-.  \]
The $e^+e^-$ pair mis-identification as a single \v{C}erenkov shower can be achieved in one of two ways: \emph{i)} events with sufficiently overlapping $e^+ e^-$ \v{C}erenkov rings such that the final state is indistinguishable from a single $e$-like event, and \emph{ii)} highly energy-asymmetric $e^+e^-$ pairs, in which one lepton is of sufficiently low energy as not to be resolved consistently.
The true percentage of ``overlapping'' and ``asymmetric'' events is hard to
compute without a dedicated experimental analysis through the reconstruction
techniques of the collaboration. In particular, MiniBooNE's detailed optical model \cite{AguilarArevalo:2008qa} and the use of electron and $\pi^0$ likelihood functions in the final selection are difficult to reproduce externally and in this analysis we simply estimate the thresholds at which we expect these distinctions to be possible. The MiniBooNE analysis employed significant efforts to remove NC $\pi^0$ events in which there were two distinct \v{C}erenkov rings. To this end, every potential event was fitted both with the single-shower ($e$) and two-shower ($\gamma \gamma$) hypothesis, and the log-likelihood ratio of these fits being used in the final selection. To help prevent cases in which the algorithm finds a better two shower fit even in the case of true electron events, an additional energy-dependent requirement that the invariant mass of the two shower candidates was much less than the pion mass was included, $m_{\gamma \gamma} < 0.3203 + 0.7417 E_\text{e}+ 0.2738 E_\text{e}^2$, where $E_\text{e}$ is the energy assuming single electron hypothesis in GeV \cite{Patterson:2007zz}. In this way even events in which one of the daughter $\pi^0$ photons was low energy could be rejected without suffering too large a reduction in signal efficiency. In the case of our $e^+e^-$ signal from a 3-body sterile neutrino decay, even if the fit slightly favored a two shower hypothesis it could still be accepted as signal if the event invariant mass does not violate this bound. 

In order to give a  quantitative estimate for the ``overlapping'' and ``asymmetric'' fractions, we refer to MiniBooNE's own analysis. Although the CCQE selection includes a cut of $>200$ hits in the main detector tank (180 hits approximately corresponding to the upper bound of a Michel electron, 52.8 MeV), this applies to the event as a whole, and provided that the most energetic shower is greater than this then it would be possible to find a significantly lower energy shower alongside it. In the final $\pi^0$ selection the lowest events which MiniBooNE successfully detected had a second shower with $\approx$ 30 MeV reconstructed energy \cite{AguilarArevalo:2008xs}, although at a low efficiency. In parallel with this, the most recent MiniBooNE data shows that the observed excess is solely contained in a bin of angular separation between electrons $\theta_\text{sep} < 16^\circ$ \cite{winecheese}. As such we take a conservative definition of an $e^+e^-$ pair to be overlapping when the true angular separation between the fermions is very small, $\theta_\text{sep} < 5^\circ$, and asymmetric events being those for which the softest particle of the $e^-e^+$ pair carries less than $30$ MeV true total energy. In both cases we also demand the $e^+e^-$ pair invariant mass to be less than the threshold utilized by the MiniBooNE CCQE selection analysis, defined above, which helps to ensure these events would not be reconstructed as a two-showered $\pi^0$ event.

The degree at which a given $e^+e^-$ pair coming from a three-body decay
(with an associated light active neutrino) meets our mis-reconstruction criteria depends predominantly on the boost factor of the parent heavy neutrino. We have studied this via a dedicated Monte Carlo simulation of decay events, confirming that the percentage of $e^+e^-$ decays in our model which are classified as asymmetric or overlapping events is mostly insensitive to the $Z^\prime$ mass, with typical values ranging from 40\% (for $m_4$ of 50 MeV) to below 10\%  (for $m_4 \geq 200 $ MeV). 
The decays which do not satisfy our conditions would appear as a diffuse background to two shower events, such as the abundant NC-induced $\pi^0 \rightarrow \gamma \gamma$ events. We have checked that most events outside our selection region have a dilepton invariant mass above $80$ MeV, a threshold used to define the MiniBooNE $\pi^0$ data sample \cite{Anderson:2010zz}. We note that MiniBooNE did observe a slight excess in NC $\pi^0$ events relative to their Monte Carlo predictions \cite{AguilarArevalo:2009ww}, although this was corrected for in the CCQE $\nu_e$ analysis.

\begin{figure*}[t]
    \begin{subfigure}{0.5\textwidth}
        \centering
	 \includegraphics[width=\linewidth]{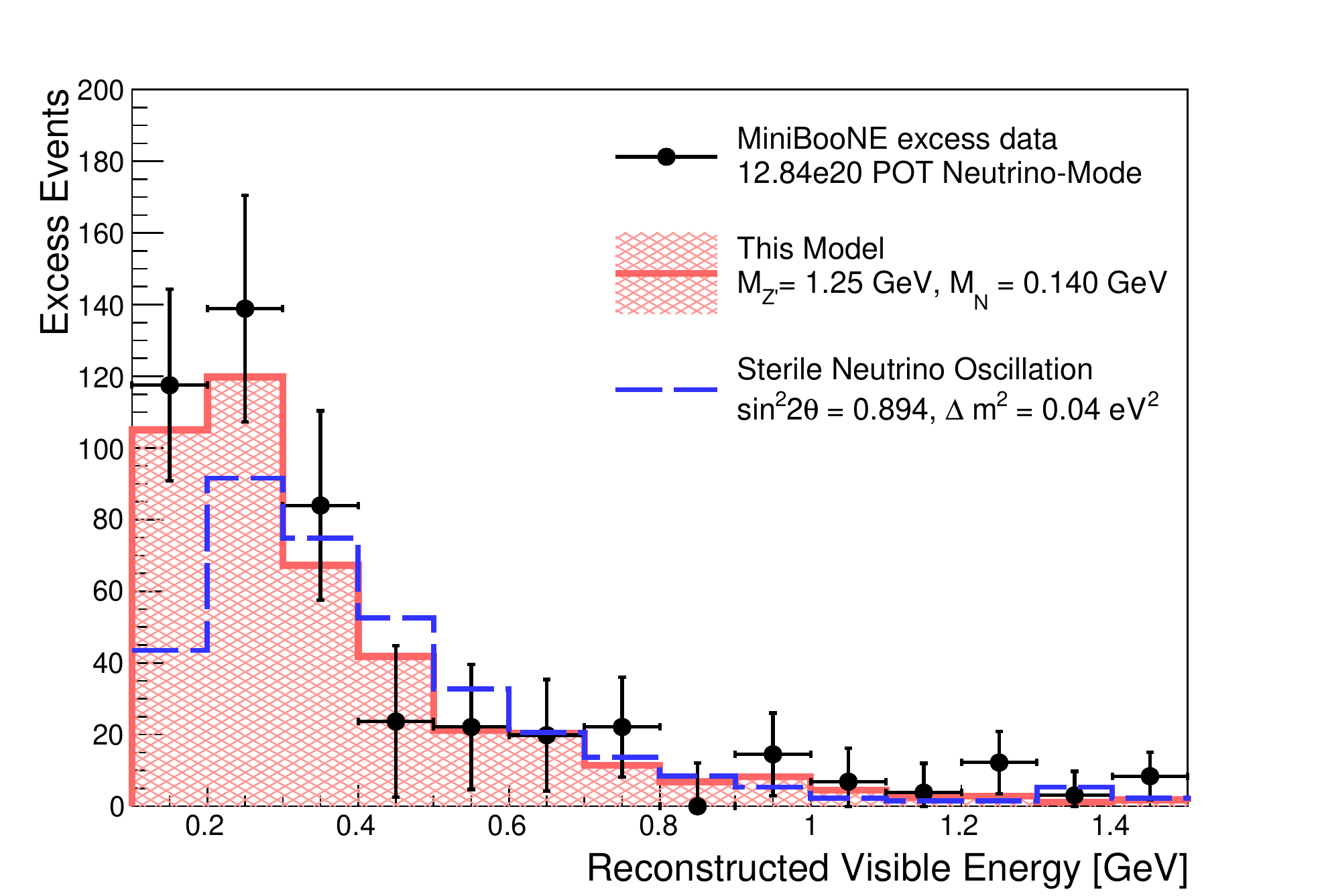}
    \end{subfigure}%
    ~ 
    \begin{subfigure}{0.5\textwidth}
        \centering
 \includegraphics[width=\linewidth]{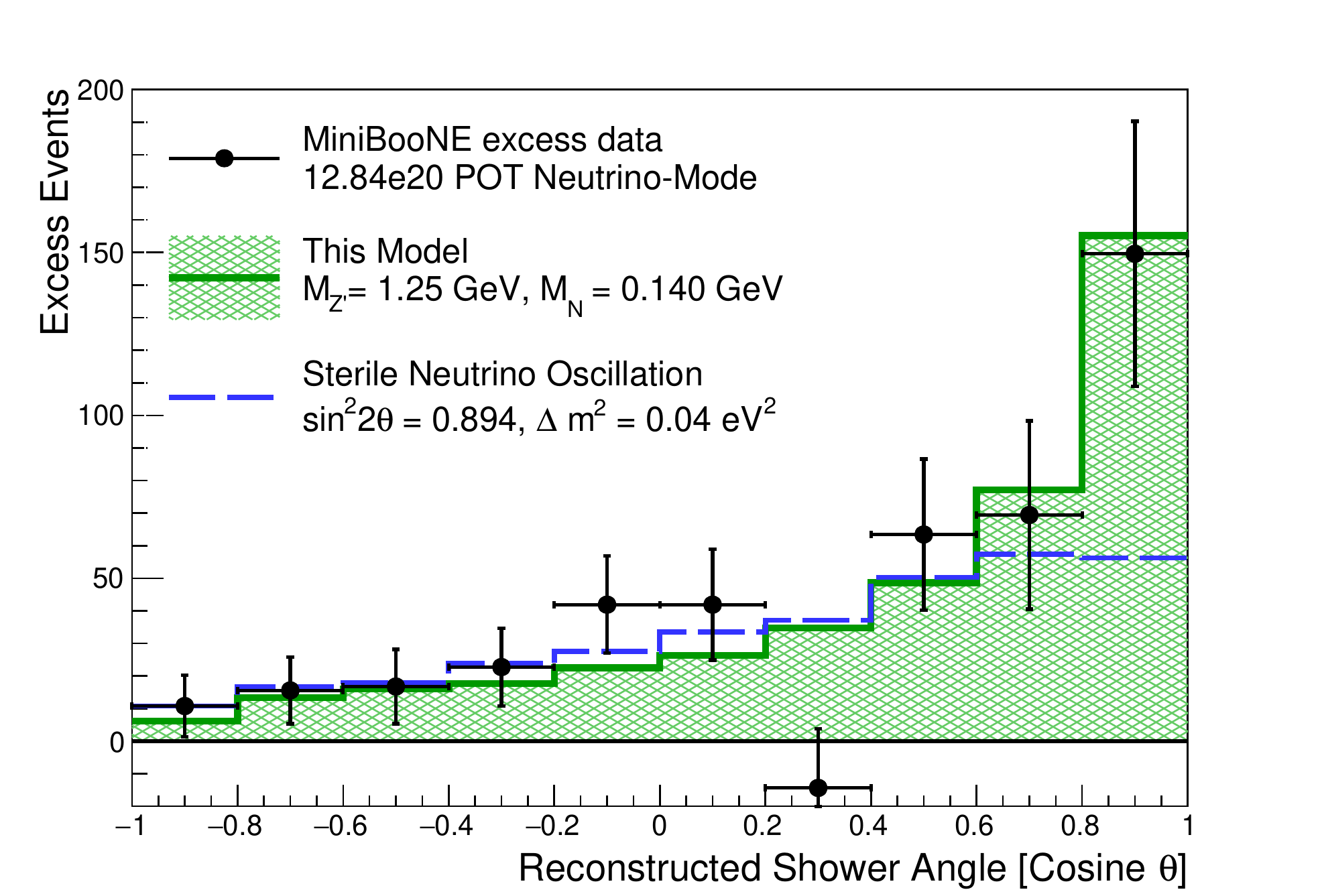}
    \end{subfigure}
\caption{\label{fig:spectra} Our model predictions in relation to the MiniBooNE excess, after subtracting predicted backgrounds, in both reconstructed visible energy (left) and reconstructed shower angle relative to the beam-line (right), for a $0.14$ GeV sterile neutrino and $1.25$ GeV $Z^\prime$. In a minimal realization, this requires neutrino mixings of $|U_{\mu 4}|^2 = 1.5\times10^{-6}$,  $|U_{\tau 4}|^2 = 7.8\times10^{-4}$ and kinetic mixing $\chi^2 = 5\times10^{-6}$, corresponding to a total decay length of 1~m. Excellent agreement is observed in both the energy and angular spectrum. MiniBooNE's best-fit sterile neutrino oscillation model is shown for comparison (blue dashed line).}
\end{figure*}

In order to identify the parameter space favoured by our explanation of the LEE, we have performed a Monte Carlo simulation of the both the scattering and subsequent decay process, obtaining the visible energy and angular distribution of $e$-like events which meet our mis-reconstruction criteria. We fully incorporate the MiniBooNE detector and selection efficiencies for the CCQE $\nu_e$ analysis, as published in the data release for \cite{Aguilar-Arevalo:2012fmn}.
We show the allowed regions of parameter space that can explain the MiniBooNE LEE in Fig.~\ref{fig:ee_cuts}.
The left (right) panel shows the result of a shape only $\Delta \chi^2$ fit to the energy (angular) distributions of the LEE \cite{Aguilar-Arevalo:2018gpe}. The angular and visible energy spectra have been fit separately, as we do not have access to the fully correlated distributions, although their strong agreement in the preferred region leads us to expect such correlations will not significantly alter our result. 
The goodness of our fit can be seen in Fig.~\ref{fig:spectra} where the predictions of a representative model is shown for both reconstructed visible energy and shower angle. This figure assumes a sterile neutrino of mass $0.14$ GeV and a $Z^\prime$ of mass $1.25$ GeV. Excellent agreement is seen in both, with the $Z^\prime$ being heavy enough to produce a sufficiently isotropic signal, and the sterile neutrino mass allowing for the correct, steeply rising, visible energy spectra that was observed. 

These figures report a shape-only analysis which mainly depends on the masses of the new particles. The total event rate is instead controlled by the specifics of the decay and by the allowed values of $\chi$ and $|U_{\alpha 4}|$, factors which can change significantly from model to model. Deferring a thorough exploration of these issues to future work, we present here a concrete minimal realization of our explanation.
As a representative value, we find that we can explain the MiniBooNE LEE with neutrino mixing angles of $|U_{\mu 4}|^2 = 1.5\times10^{-6}$ and $|U_{\tau 4}|^2 = 7.8\times10^{-4}$, a kinetic mixing strength of $\chi^2 = 5\times10^{-6}$ and a coupling of $g^\prime=1$. In this case, the hierarchy in mixing angles leads to a dominant visible decay of $\nu_\tau e^+e^-$ with a total decay length of $\mathcal{O}(1)$~m. We find an expected event rate of 430 LEE events produced from scattering inside the detector. 

Finally, we note that our estimates are based only on production from scattering inside the detector. As the MiniBooNE analysis does not rely on the reconstruction of the scattering vertex, the potential exists for additional dirt events to contribute to our signal. These are expected to have the same kinematic properties as those simulated above, and will generally increase our event rates, moving our estimated values of $\chi^2$ and $|U_{\alpha 4}|^2$ to lower values. 

%


\subsection{\label{sec:bounds} Constraints on a minimal realization}

In this section we will show that our minimal realization based on a hierarchical mixing pattern can produce a sufficient rate of signal events whilst satisfying all current bounds. We stress that this is just one concrete way to achieve the necessarily small decay lengths that are required. For instance, if there are additional hidden sector particles charged under $U(1)^\prime$ with which the sterile neutrino and $Z^\prime$ can interact with, these could lead to rapid $\nu_4$ decays. A dedicated study is required to understand the bounds on such models or other variants and is beyond the scope of the current work. 

Both the kinetic mixing, $\chi^2$, and active-sterile mixing elements, $|U_{\alpha 4}|^4$, have been bounded by many experiments in the past. However, the assumptions used to convert the non-observation of a signature into constraints on the fundamental parameters are not universally applicable in non-minimal models. By having at the same time heavy neutrinos and a $Z^\prime$, the separate bounds on these particles need to be revised and in some instances they become significantly weaker than in the standard case, so that values required by our minimal realization to explain the MiniBooNE excess are allowed.

Specifically, in our model, there are two subtleties in the interpretation of published 
bounds: we have \emph{i)} large invisible and nearly-invisible branching ratios for the $Z^\prime$ and $\emph{ii)}$ an unstable heavy neutrino which decays within distances of the order $1$--$10$~m. 
Any experiment which looked for the visible decays of on-shell $Z^\prime$
particles must be reconsidered taking into account the invisible decays of the new boson which dominate and lead to a visible branching fraction which is suppressed by a factor of $\chi^2$. Published bounds subject to this weakening include BABAR \cite{Lees:2014xha}, KLOE \cite{Archilli2012251} and A1/MAMI \cite{Merkel:2014avp}, which searched for final state electrons, by approximately a factor of a $\chi$. For a $Z^\prime$ of mass $1.25$ GeV the most stringent bound, BABAR, becomes $\chi^2 \leq 7\times 10^{-4}$ at the 90\% C.L, allowing for kinetic mixings sufficiently large to produce enough events as required in Fig.~\ref{fig:spectra}.  

The semi-invisible decays of the $Z^\prime$ might be observable in some cases, and offer a novel means to test this model. Any sterile neutrino in the final state will decay over distances of the order of few meters. The 3-body heavy neutrino decay will produce multi-lepton final states with a broad invariant mass spectrum due to final state neutrinos, and would not pass the peak-hunting cuts of \emph{e.g.} $4$-lepton searches at BABAR. A promising route to test these events would be a search for the displaced vertices of these secondary decays. A $Z^\prime\to \nu_4\nu_4$ decay could be observed as two dilepton pairs from displaced vertices, neither pointing to a common origin (as each is associated with a three-body decay with missing neutrinos). To the best of our knowledge, this would not have been picked up in existing analyses but suggests an interesting means of testing models of this kind.

Experimental bounds on possible active-heavy neutrino mixing will also be affected by
the new interactions. Enhanced decay rates of the sterile neutrino would naively increase the sensitivity of beam dump experiments, \eg\ PS191 \cite{Bernardi:1986ny} or NuTeV \cite{Vaitaitis:1999wq}. However, once the rate increases sufficiently, heavy sterile neutrinos produced in the beam will decay before reaching the detector and the bounds will be removed. This greatly restricts the applicability of published constraints arising from such beam dump experiments, where the flux of new heavy states is assumed to be suppressed by mixing alone. For the parameters of our minimal realization, the sterile neutrino has a decay length of around 1~m, severely weakening the bounds for $U_{\mu 4}$ set by PS191 (at a baseline of 128m) and also $U_{\tau 4}$ set by experiments such as NOMAD\cite{Astier:2001ck} (835m) and CHARM \cite{Bergsma:1985is} (487m).

Peak search experiments, which look for resonant bumps in the associated
leptons of meson decay, \eg\ $K^\pm\rightarrow \mu^\pm \nu_4$, have no dependence on the subsequent decay rate and fully apply in our model \cite{Kusenko:2004qc}. The value of $|U_{\mu 4}|^2$ required by our benchmark point is $1.5\times10^{-6}$, below the bounds of $2.3\times10^{-6}$ coming from the kaon peak search for a 140 MeV sterile neutrino.

Neutrino trident production could also place bounds on our new mediator \cite{Altmannshofer:2014pba}. Although, thanks to the significantly suppressed active-active-$Z^\prime$ vertex, these contributions are negligible. Previous probes of neutrino trident production such as CHARM and CCFR would still potentially be sensitive to new contributions if a sterile was produced in the final state. However, depending on the rate and daughters of any subsequent decay of the heavy neutrino, events may not have been included in the SM neutrino trident search. 
We note that a decaying heavy sterile produced in our model would look identical to the SM electron trident process: $\nu_\mu {\cal N} \rightarrow \nu_\mu  e^+ e^- {\cal N}$. This process has yet to be observed in the SM due to the complexity of resolving the final state, but is expected to be observable by upcoming neutrino beams such as DUNE \cite{Ballett:2018uuc}, which will further constrain our model.

Although we leave an exhaustive scan of the whole parameter space to a future study, we have shown a benchmark point in a minimal realization of our model, whose angular and energy spectra are in excellent agreement with the MiniBooNE LEE and which provides a satisfactory explanation with parameters allowed by the current experimental bounds.

\section{Summary and Predictions for MicroBooNE}
We have discussed a novel explanation of the MiniBooNE low-energy excess based
on heavy sterile neutrino production and decay inside the detector, both of
which are mediated by a novel $Z^\prime$.  The explanation hinges on the
mis-identification of the EM shower induced by a combination of highly asymmetric and overlapping $e^+e^-$ pairs, which we argue happens for a sufficient fraction of decays in the sterile neutrino and $Z^\prime$ mass regions of interest. 
We have shown a specific phenomenological model based on a sterile neutrino
coupling to a hidden-sector $U(1)^\prime$, which provides a remarkable fit to both the energy and angular spectra of the LEE. This spectral agreement favours sterile masses of $100\lesssim m_4 \lesssim 250$ MeV and $Z^\prime$ masses above 1 GeV.
We have stressed that the event rate itself is dependent on the other parameters and on specific assumptions made on the model.
We have presented a specific realization requiring no additional particles beyond the sterile and $Z^\prime$, but with a hierarchy in the sterile-active mixing angles. This model can produce the correct number of events while maintaining the excellent spectral agreement with the LEE as in our more general model. Interestingly, the interplay between $U(1)^\prime$ kinetic mixing and neutrino mass mixing leads to crucial model-dependent reinterpretations of the bounds on conventional $\mathcal{O}$(100)~MeV heavy neutrinos so that the values of the parameters required to explain the MiniBooNE LEE are allowed.
We emphasize that this is only a minimal realization, and there are many variants on the core mechanism of this interpretation. For example, the production and decay could proceed via different light mediators, or the decay rate could be enhanced in a model with two sterile neutrinos where the heavier subsequently decays into the lighter one, as can arise in linear/extended see-saw models.

Our explanation can be falsified at contemporary short-baseline experiments such as MicroBooNE, whose liquid argon technology will crucially have access to topological and calorimetric means to distinguish electromagnetic showers of electrons from those of photons. We estimate that MicroBooNE would see around $150~(75)$ LEE signal-like events in the planned 6.6e20 POT exposure, assuming a selection and reconstruction efficiency of 80\% (40\%) \cite{Antonello:2015lea}. These signal-like events are highly asymmetric or overlapping and would be split between MicroBooNE's $\gamma$-like LEE search (overlapping) and their $e$-like search (asymmetric). The model presented here has the unique feature that accompanying the 150 LEE events will be around 500 $e^+e^-$ pairs with more easily distinguishable showers\footnote{Due to the higher atomic mass of Argon, this would be further enhanced by an proportional increase in coherent events, although these additional events would favour forward-going and overlapping electron pairs.}.
Unlike at MiniBooNE, where the two-shower events are competing against the large sample of NC $\pi^0$ events, MicroBooNE would see a novel signal of two $e$-like showers originating from a single vertex with no other associated hadronic activity. The only other common interactions that produces two electromagnetic showers in MicroBooNE are NC $\pi^0$ events in which both photons convert to $e^+e^-$ pairs within a centimetre in conjunction with failing the electron-photon separation $dE/dx$ calorimetric cuts \cite{Acciarri:2016sli}. 
As such we believe this is an extremely clean signal channel, allowing for the direct test of this class of models at MicroBooNE.\\

\emph{{\bf Note added:} During the very final stages of this work, an explanation~\cite{Bertuzzo:2018itn} for the MiniBooNE LEE appeared which also invokes a light $Z'$ and a sterile neutrino. The latter is produced in $Z^\prime$-enhanced NC interactions in the detector and subsequently decays into a light neutrino and an on-shell $Z'$ which itself decays rapidly into an $e^+e^-$ pair. The signature is two strongly-overlapping electrons. Although such explanation can also be achieved in our model if $m_{Z'} < m_4$, we focus here on the alternative case of $m_{Z'} > m_4$ and defer a more in-depth analysis of other variants of the explanation to future work. }\\

\begin{acknowledgments}
We would like to thank M.~Hostert, J.~J\"ackel, A.~Szelc, J.~Asaadi and M.~Shaevitz for discussions 
on various aspects of $Z^\prime$ phenomenology and detector functionality.

This work has been supported by the European Research Council under ERC Grant ``NuMass'' (FP7-IDEAS-ERC ERC-CG 617143), by the European Union Horizon2020 ELUSIVES ITN (H2020-MSCA-ITN-2015, GA-2015-674896-ELUSIVES), and InvisiblesPlus RISE (H2020-MSCARISE-2015, GA-2015-690575-InvisiblesPlus). SP gratefully acknowledges partial support from the Wolfson Foundation and the Royal Society.
\end{acknowledgments}


\bibliographystyle{apsrev4-1}
\bibliography{main}{}

\end{document}